\documentclass[aps,prb,twocolumn,groupedaddress,showpacs]{revtex4}

\usepackage{subfigure}
\usepackage{graphicx}
\usepackage{bm}
\usepackage{amsmath}
\usepackage{epstopdf}
\usepackage{dcolumn}

\newcommand{\STO}{SrTiO$_3$}
\newcommand{\LAO}{LaAlO$_3$}

\newcommand{\etal}{\emph{et al.}}

\begin{document}

\title{Tunneling into a quantum confinement created by a single-step nano-lithography of conducting oxide interfaces}

\author{E.Maniv}
\affiliation{Raymond and Beverly Sackler School of Physics and Astronomy, Tel-Aviv University, Tel Aviv, 69978, Israel}
\author{A.Ron}
\affiliation{Raymond and Beverly Sackler School of Physics and Astronomy, Tel-Aviv University, Tel Aviv, 69978, Israel}
\author{M.Goldstein}
\affiliation{Raymond and Beverly Sackler School of Physics and Astronomy, Tel-Aviv University, Tel Aviv, 69978, Israel}
\author{A.Palevski}
\affiliation{Raymond and Beverly Sackler School of Physics and Astronomy, Tel-Aviv University, Tel Aviv, 69978, Israel}
\author{Y.Dagan}
\affiliation{Raymond and Beverly Sackler School of Physics and Astronomy, Tel-Aviv University, Tel Aviv, 69978, Israel}

\begin{abstract}
A new nano-lithography technique compatible with conducting oxide interfaces, which requires a single lithographic step with no additional amorphous layer deposition or etching, is presented.
It is demonstrated on \STO/\LAO~interface where a constriction is patterned in the electron liquid.
We find that an additional back-gating can further confine the electron liquid into an isolated island.
Conductance and differential conductance measurements show resonant tunneling through the island.
The data at various temperatures and magnetic fields are analyzed and the effective island size is found to be of the order of 10nm. The magnetic field dependence suggests absence of spin degeneracy in the island. Our method is suitable for creating superconducting and oxide-interface based electronic devices.
\end{abstract}

\pacs{81.07.-b,73.23.-b, 73.20.-r}

\maketitle
\section{INTRODUCTION}
When free electrons are confined into a structure whose dimensions are smaller than their Fermi wavelength, their energy states become quantized. The resulting electronic level spacing increases as the dimensions of the confinement region decrease. Gate voltage can tune the chemical potential to match between the discreet spectrum of the confinement and the continuum in the reservoir realized by the leads.
Such matching or resonance occurs when the electronic population in the confinement region is changed by a single electron. In addition to the level spacing, adding a single electron to the confinement requires surmounting the charging energy $E_C=\frac{e^2}{2C}$ with $C$ the capacitance. This energy is usually larger than the spacing between the levels in the confinement, for most of metallic and semiconducting dots. Such single electron transistors (SETs) has been realized in many semiconductor devices \cite{kastner1992single}.
\par
While in semiconductor devices unconfined electrons can be described in the non-interacting particles picture, two dimensional electron gas formed at the interface between the insulating oxides \STO~and \LAO~\cite{ohtomo2004high} can undergo a variety of phase transitions \cite{thiel2006tunable, reyren2007superconducting, bert2011direct, ron2014anomalous} and electronic correlations should be taken into account \cite{breitschaft2010two,maniv2015strong}. It has been shown that back gating can be very efficient for carrier modulation in mesoscopic samples on \STO~as a result of field focusing by the large dielectric constant \cite{rakhmilevitch2013anomalous}. However, nano-lithography in \STO~based interfaces is challenging due to its sensitivity to vacancies and impurities resulting in spurious parallel conductivity \cite{kalabukhov2007effect}.
\par
In order to define a conducting channel for \STO/\LAO~interface usually an amorphous hard mask \cite{schneider2006microlithography} is used. This process requires more than one deposition step \cite{goswami2015quantum}. Another option to define nanostructures is using conducting atomic force microscope (AFM) tip biased against the \STO/\LAO~interface below its conductivity threshold of four unit cells \cite{cen2009oxide}. This method has proven to be very useful for producing nanometric devices, in particular a SET \cite{Levy2015QD}. However neither back gate nor top gate can be used and the resulting devices are sensitive to temperature cycles and electric fields.
\par
\begin{figure}
\begin{center}
\includegraphics[width=1\hsize]{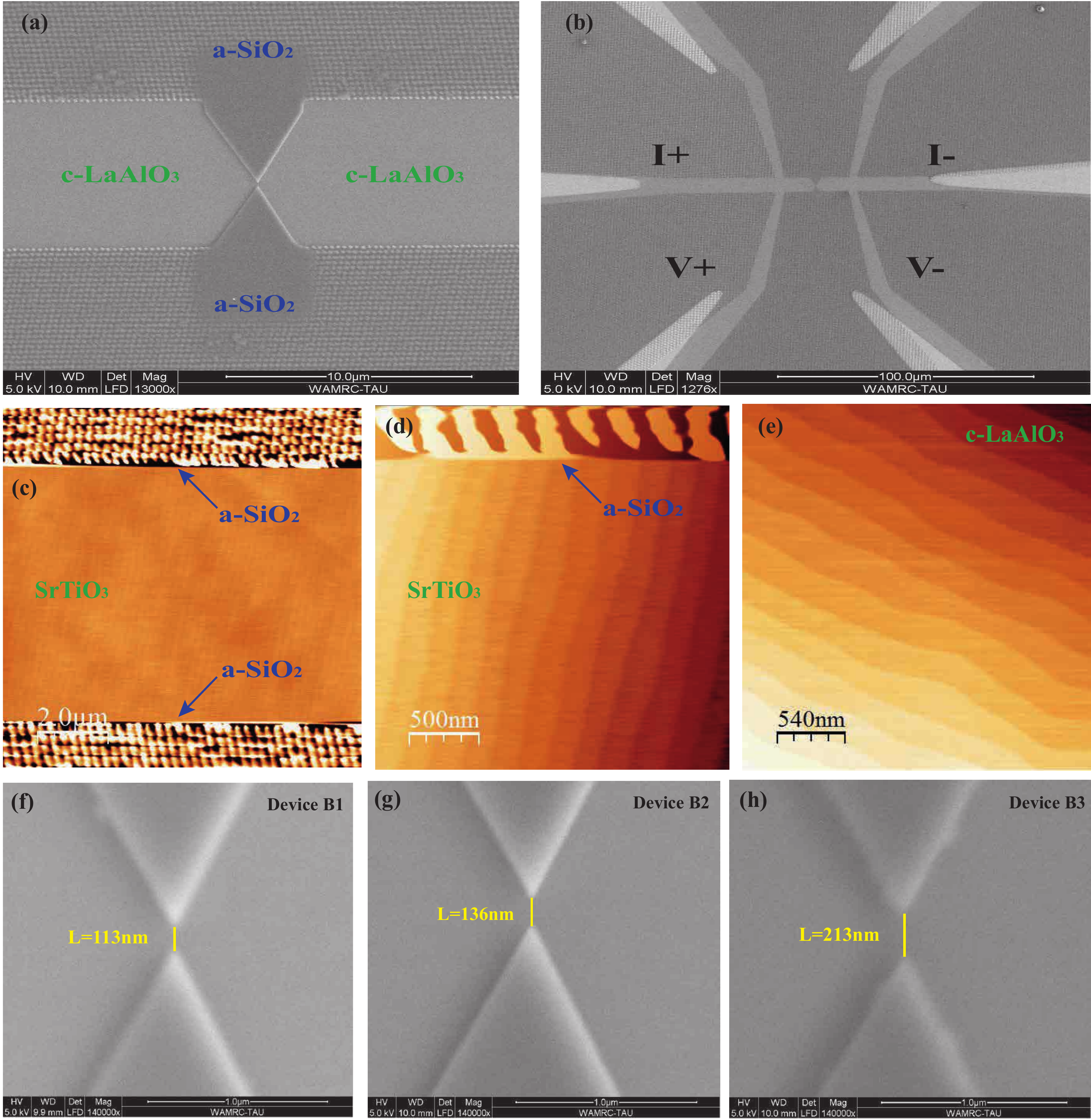}
\caption{(color online). (a) SEM image of the HSQ electron-beam-lithography pattern of sample B2. The two different regions (conducting crystalline and amorphus insulating) are marked (c-\LAO~ and a-SiO$_2$). (b) Overview SEM image showing the device structure and contact configuration for sample B2. (c)-(d) AFM images of a patterned HSQ hard mask, which corresponds to the c-\LAO~ region in (a) (image taken before \LAO~deposition). The sharp contrast between the a-SiO$_2$ and the atomically smooth \STO~is conspicuous. No traces of the a-SiO$_2$ are visible in the \STO~regions. (e) AFM image of one of the c-\LAO~ contacts (before ion milling). A clear step-terrace structure is observed compatible with the \LAO~unit cell. (f)-(h) SEM images focused on the constriction regions of samples B1, B2 and B3; their sizes are marked by yellow lines.}
\label{Sample Characterization}
\end{center}
\end{figure}
Here we demonstrate a new and simple lithography technique applicable for conducting oxide heterostructures, which requires only a single step of lithography and deposition. The hard mask can be defined with a resolution of sub-10nm \cite{grigorescu2007sub}. Applying this method we successfully fabricated narrow quantum constrictions resulting in quantum dot (QD) formation inside the constriction and performed the electron-level spectroscopy of the dot.
\par
\begin{figure}
\begin{center}
\includegraphics[width=1\hsize]{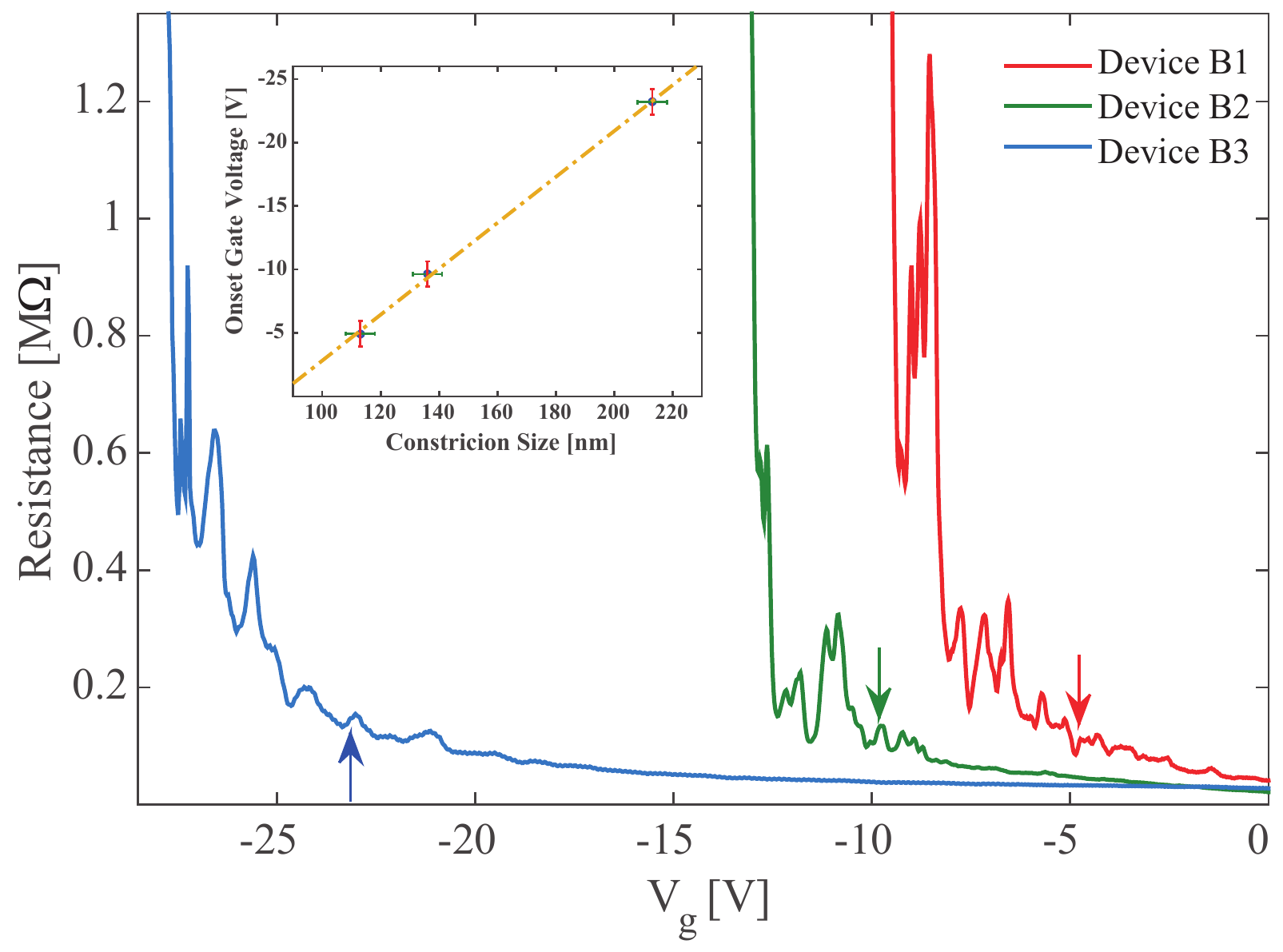}
\caption{(color online). Resistance versus gate voltage sweeps measured at 1.6K for samples B1-B3, whose images are presented at Figure\ref{Sample Characterization}(f)-(h). Arrows depict the gate voltage where the dip-peak structure onsets. The inset shows the extracted onset gate voltage versus the nominal constriction size. The line is a guide to the eye.}
\label{SizeVsGate}
\end{center}
\end{figure}
\par
\section{METHODS}
In order to define a nanometric constriction we used a new lithography approach. We spin-coated atomically flat TiO$_2$ terminated \STO~substrates with 140nm of hydrogen silsesquioxane (HSQ). The devices were then defined using e-beam lithography. This results in a hard mask of amorphous SiO$_2$ (a-SiO$_2$) on top of the \STO. Then 10 unit cells of epitaxial \LAO~were deposited using pulsed laser deposition as described in reference \cite{shalom2009anisotropic}. Gold gate electrodes were evaporated to cover the back of the substrate. The leakage current was unmeasurably small ($<1 pA$) in the entire range of applied voltages. In some samples, gate voltage range under study varied from one cool-down to another. Conductance and differential conductance were measured using a lock-in amplifier-based technique.
\par
\section{RESULTS AND DISCUSSION}
\subsection{Characterization of the devices}
In Figure \ref{Sample Characterization}(a) we show scanning electron microscope (SEM) image of one of the defined structures (sample B2). Electrical conductance appears only in \STO~regions under the crystalline \LAO~(c-\LAO) while the regions under the a-SiO$_2$ are insulating. An overall view of the device is shown in Figure \ref{Sample Characterization}(b); the electrical circuit connection configuration is noted. Ti/Au pads are deposited after ion milling through the \LAO~layer to form ohmic contacts with the conducting interface between the c-\LAO~ and \STO~(Figure \ref{Sample Characterization}(b)). In Figure \ref{Sample Characterization}(c) we show an AFM image of a hard HSQ mask on \STO. The HSQ was removed from regions unexposed to the electron beam. Figure \ref{Sample Characterization}(d) focuses on the vicinity of the \STO-HSQ boundary. From both Figures \ref{Sample Characterization}(c) and \ref{Sample Characterization}(d) it is clear that the HSQ is completely removed from the desired regions leaving an atomically flat \STO~surface with the typical step-terrace morphology. In Figure \ref{Sample Characterization}(e) we show an AFM image of the c-\LAO~ region. The step-terrace morphology is indicative of atomically flat and crystalline \LAO~layer. In Figure \ref{Sample Characterization}(f)-(h) we show constrictions with increasing sizes demonstrating our capability to control the device dimensions.
\par
In Figure \ref{SizeVsGate} we show the resistance versus gate voltage data taken at 1.6K for samples B1-B3 whose corresponding images are shown in Figure\ref{Sample Characterization}(h)-(f). As expected the resistance is increasing with negative gate voltage but an unusual pattern of dips and peaks appears at a certain onset voltage. This onset increases monotonically with the size of the constriction as shown in the inset of Figure\ref{SizeVsGate}. For sample A similar gate scans are shown in the supplementary part of the paper (Figure S1~\cite{SuppQD}). In sample A for constriction size of 370nm we could not see the dip-peak structure with gate voltages of up to -60Volts. The geometry of these five devices defer from each other only by the constriction size (samples A and B were made on different substrates). Namely, they have the same bridge width and feature angles. We further note that samples of similar size that lack the sharp-edge-constriction geometry exhibit resistance, which is monotonic with back-gate voltage, suggesting that the effect we observe is not related to intrinsic inhomogeneities due to disorder over a length scale of a few 10nm (See Figure S2~\cite{SuppQD}).
\par
In conclusion, we conjecture that the dip-peak pattern is a result of tunneling through an isolated conducting island formed in the constriction. The necessary conditions for the island formation are: sharp features and a narrow constriction.
\par
\begin{figure}
\begin{center}
\includegraphics[width=1\hsize]{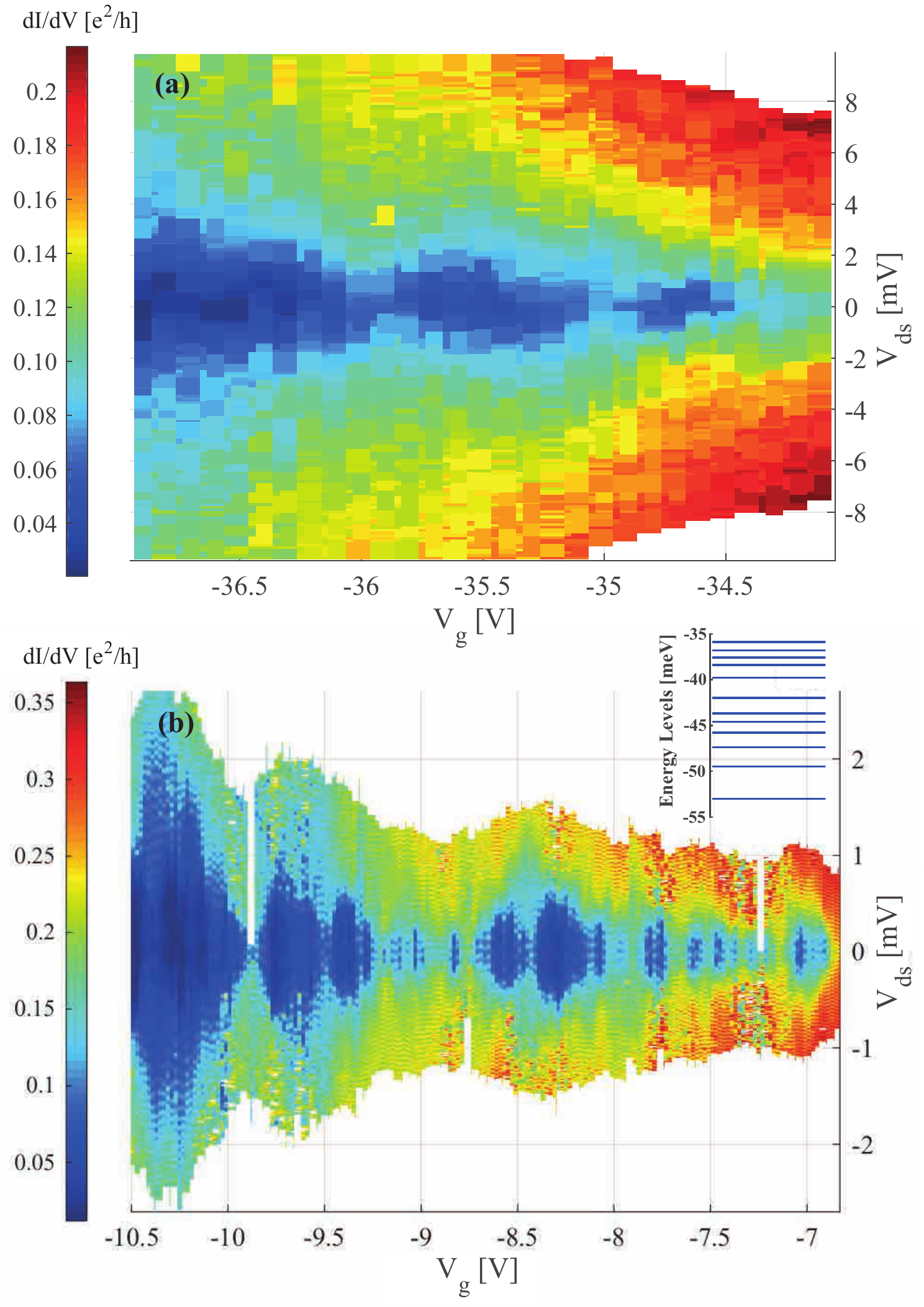}
\caption{(color online). (a) and (b) Color coded differential conductance (dI/dV$_{ds}$) of sample A1 and sample B2 at 1.5K and 160mK, respectively, as a function of both the bias voltage V$_{ds}$ and the back gate voltage V$_g$. Resistive diamond regions (blue) are seen in both samples. Note that color code for (b) is re-scaled for clarity compared to (a). We plot the matching energy levels of sample B2 in the inset of (b) using the calculated conversion factor $\alpha$. The variations in the level spacing are the usual ones typically observe for a generic quantum dot in the presence of impurities and/or irregular boundary.}
\label{Diamonds}
\end{center}
\end{figure}
\subsection{Quantum-dot characteristics and analysis}
It is possible to accumulate electrons in the island using either back gate (V$_g$) or drain-source voltage (V$_{ds}$). Scanning both gate-voltages should result in diamond-shape regions of low conductance in V$_g$-V$_{ds}$ diagram. Such a plot is shown in Figure \ref{Diamonds}(a) for sample A1 at 1.5K and in Figure \ref{Diamonds}(b) for sample B2 at 160mK.
\par
The reservoir is coupled capacitively to the island through two main channels whose capacitances are: C$_l$ and C$_{bg}$ for the leads and the back gate, respectively. The slope of the diamonds is therefore $\alpha\simeq\frac{C_{bg}}{C_l+C_{bg}}$.
From both diamond plots (Figures \ref{Diamonds}(a), \ref{Diamonds}(b)) we estimate $\alpha$ to range between 0.004 and 0.006.  The variations of $\alpha$ with back gate voltages can be attributed to the changing dielectric constant of the \STO~\cite{fuchs1999high} and the resulting change of the self capacitance of the island.
\par
The extremely large value of the dielectric constant $(\sim 24,000)$  \cite{muller1979srti} makes the \STO~ material very unique in the context of tunneling through the confined region, since the charging energy (Coulomb blockade) is expected to be suppressed by over three orders of magnitude relative to the devices of similar sizes made of metals and most semiconductors. $E_C$ is therefore negligible compared to the level spacing for such confined regions with dimensions of up to 1 micron. Consequently the Coulomb blockade can be ignored, as mentioned. Indeed, $C=8\epsilon_0\epsilon_rR$ is the self capacitance of a disk of radius R, and estimating the level spacing to be $\Delta\varepsilon=\frac{\hbar^{2}}{m^{*}R^{2}}$ for free electrons in a two dimensional confinement, where $m^{*}=0.7m_e$ is the effective mass of the lower band of \STO/\LAO~\cite{santander2011two} (for the low carrier concentration studied we expect only the lowest band to be populated) we obtain that E$_C=\delta\epsilon$ when R=30 micron. Therefore, $E_C<<\Delta\varepsilon$ for all possible island sizes within our constriction limited to few hundred nanometers. This implies that our measurement is a direct probe of the density of states of the island $\nu(\varepsilon)$. Similar arguments were given by Cheng \etal~\cite{Levy2015QD}.
\par
In an ideal QD the diamond shaped regions with vanishing conductance should be separated by sharp conductance peaks. In our case the entire picture is shifted by a conducting background and the features are asymmetric and somewhat broadened. It is possible that the islands in the various samples cannot be pictured as an isolated single QD. Rather it is likely that another non-resonant channel exists.
\par
\subsection{Fano-type behavior and temperature dependence}
Whenever resonant and non-resonant scattering paths interfere, an asymmetric Fano line shape should appear. Fano resonances have been observed in a wide range of experiments including GaAs/AlGaAs QD's \cite{gores2000fano,johnson2004coulomb}. Taking the Fano scattering formula and treating it within the Landauer-B\"{u}ttiker formalism, G\"{o}res \etal~obtained an expression, which describes the most general case of coexistence of resonant and non-resonant channels:
\begin{equation}\label{FanoEq}
G=G_{inc}+G_{0}\frac{(\xi+q)^2}{\xi^2+1}
\end{equation}
where $\xi=(\varepsilon-\varepsilon(0))/(\Gamma/2)$ is the dimensionless parameter becoming larger when the energy $\varepsilon$ is shifted away from the resonant one $\varepsilon(0)$. $\Gamma$ is the width of the resonant feature, q is the asymmetry parameter and $G_{inc}$ denotes an incoherent contribution to the conductance.
\par
\begin{figure}
\begin{center}
\includegraphics[width=0.95\hsize]{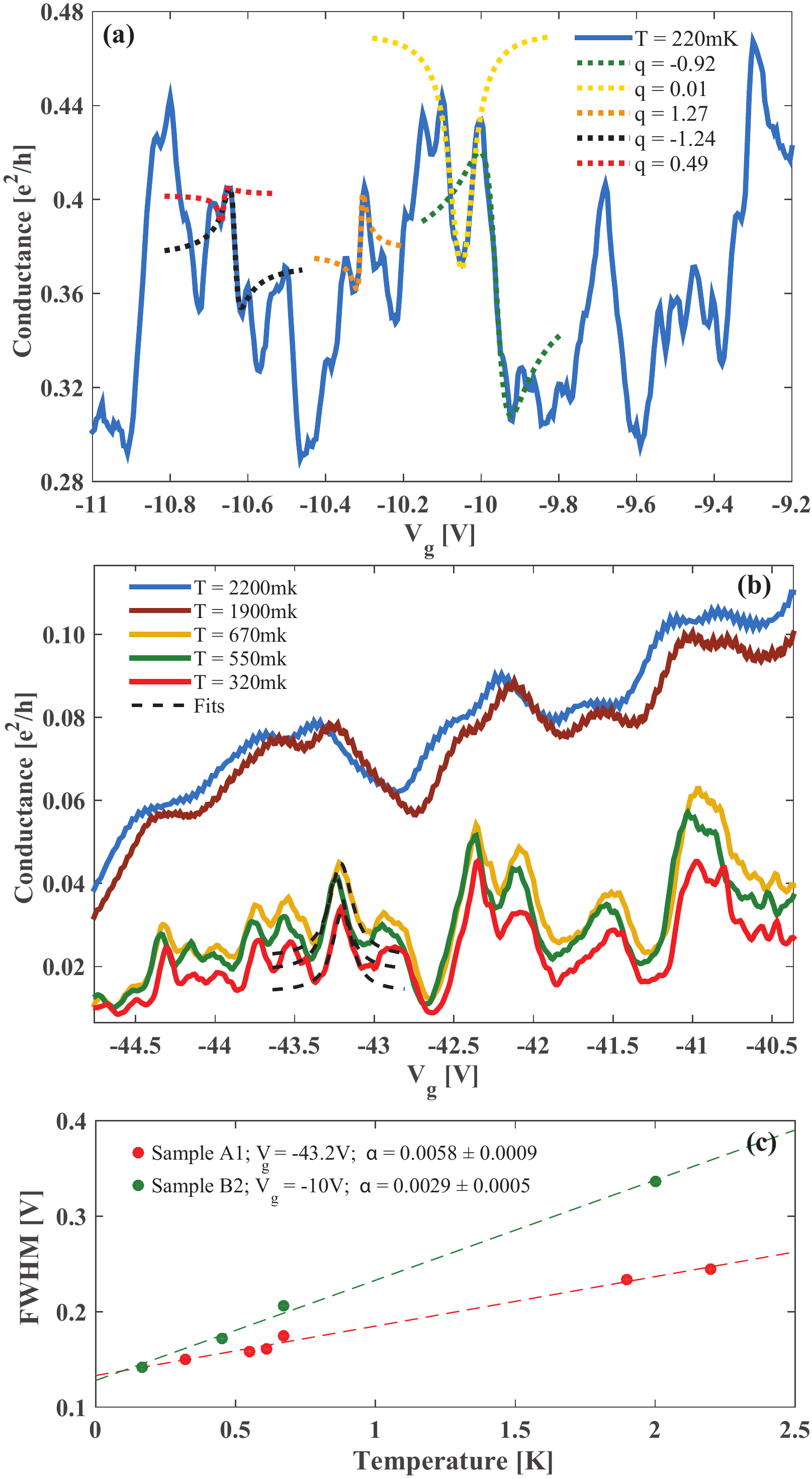}
\caption{(color online). (a) Conductance versus V$_g$ for sample B1 at 220mK. Five different fits to Eq.\ref{FanoEq} are shown. The asymmetry-parameter q is noted for each fit. (b) Conductance versus V$_g$ for sample A1 measured at various temperatures. Three fits to Eq.\ref{FanoEq} in the special case of a symmetric Breit-Wigner peak are presented for a particular V$_g=-43.2V$. (c) FWHM's extracted from the fits in (b) and similar ones, as a function of temperature for two different gate voltages for sample A1 (Figure \ref{Temperature}(b)) and sample B2 (See also Figure S3~\cite{SuppQD}). The dashed lines are linear fits. The resulting $\alpha$ factor (see text) is shown for each fit.}
\label{Temperature}
\end{center}
\end{figure}
In Figure \ref{Temperature}(a) we show the conductance through the island of sample B1 measured at 220mK. Different asymmetric dips and peaks can be observed. Fitting each dip/peak to the Fano formula (Eq.\ref{FanoEq}) gives various asymmetry parameters q, which correspond to the phase shift of the resonant signal with respect to the non-resonant one.
\par
In Figure \ref{Temperature}(b) we show the conductance through the island of sample A1 at various temperatures. The width of the peaks and the dips is reduced with lowering the temperature. In addition, more features appear below $\simeq 1K$.

By fitting the data of sample A1 at various temperatures (Figure \ref{Temperature}(b)) and sample B2 (See Figure S3~\cite{SuppQD}) to Eq.\ref{FanoEq} in the special case of a symmetric Breit-Wigner peak, which corresponds to the Fano formula in the limit $q\rightarrow\infty$ and $G_0\rightarrow0$, we can extract the width at half maxima of the resonance features (FWHM). In Figure \ref{Temperature}(b) we show examples for three such fits at V$_g=-43.2V$. In Figure \ref{Temperature}(c) we show the extracted FWHM's for sample A1 and sample B2.
\par
If the temperature is not too low ($k_{B}T>\Gamma/20$) one expects the FWHM of an electronic Lorentzian-type feature to be broadened as $\simeq3.5k_BT$ due to the temperature dependence of the Fermi-Dirac distribution. The measured FWHM in units of the gate voltage is related to the electronic energy by the conversion factor $\alpha$. The linear fits and the resulting $\alpha$'s are presented in Figure \ref{Temperature}(c). The average $\alpha=0.0044\pm0.0005$ is in reasonable agreement with the estimated value obtained from Figure \ref{Diamonds}.
\par
\begin{figure}
\begin{center}
\includegraphics[width=1\hsize]{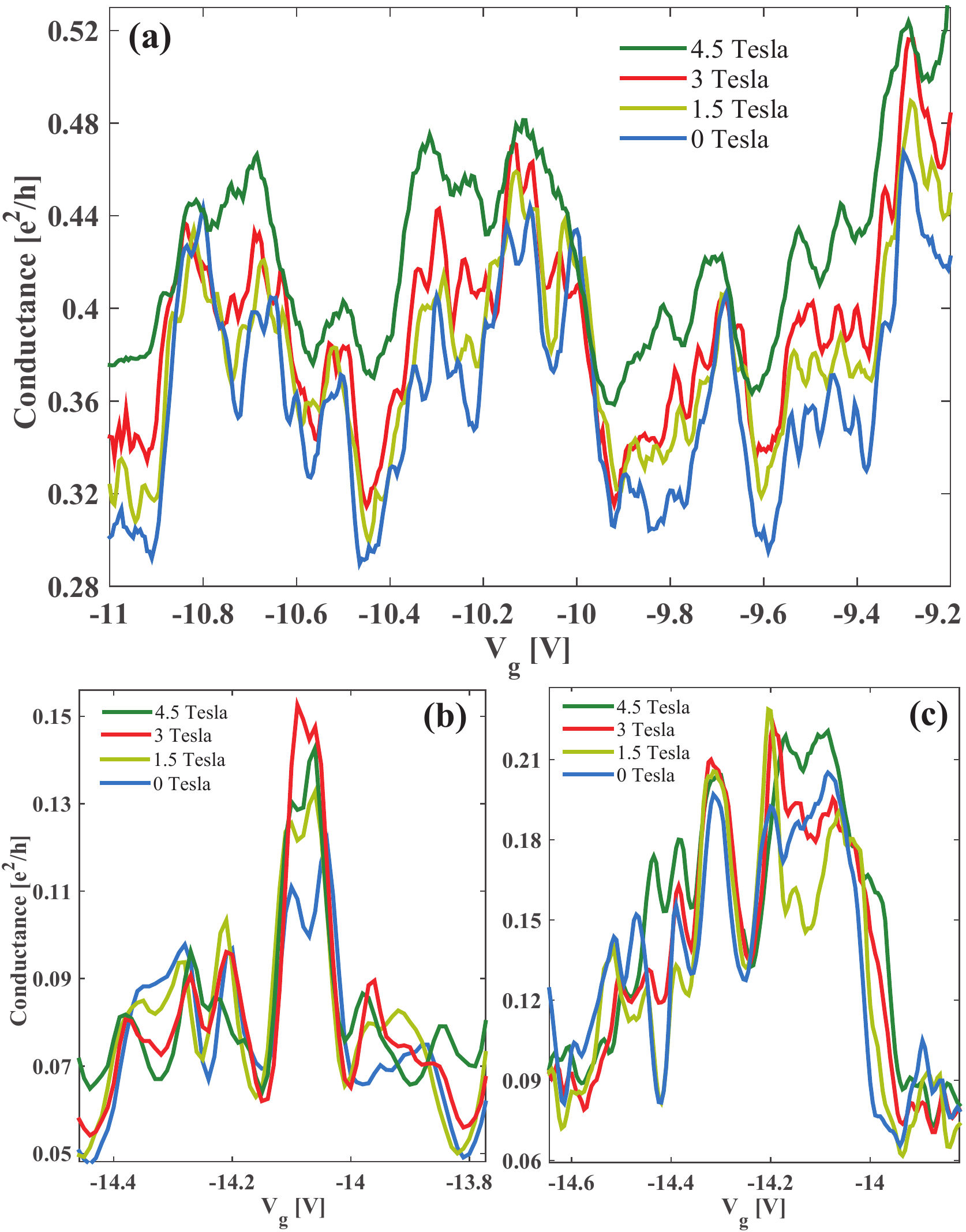}
\caption{(color online). (a) Conductance versus V$_g$ for various magnetic fields applied perpendicular to the interface for sample B1 at 220mK. (b) and (c) Conductance versus V$_g$ for sample B2 for various magnetic fields applied perpendicular to the interface at 220mK (b) and perpendicular to the current (transverse) at 160mK (c).}
\label{FanoField}
\end{center}
\end{figure}
\subsection{Level spacing calculation and discussion of dot size}
Now we can calculate the level spacing ($\Delta\varepsilon$) using the voltage difference between conductance peaks and the conversion factor $\alpha$. $\Delta\varepsilon$ varies with gate voltage, possibly due to decreasing dielectric constant. We estimate $\Delta\varepsilon$ to be between 1meV to 5meV yielding an estimated island size of 5nm to 10nm. This size estimation further supports our statement that the charging energy is negligible compared to the level spacing. The island size is different than the geometrical size of the constriction.
\par
What is the reason for the order of magnitude difference between the obtained dot size (10nm) and the lithographic dimension of the constriction (~100nm)? We have previously shown that the efficiency of the back gate improves significantly when the typical size of the sample becomes smaller than the screening length \cite{rakhmilevitch2013anomalous, maniv2015strong}. For the typical size of our constriction, the geometry of sharp edges and the large nonlinear dielectric constant we expect stray fields to become important. We conjecture that this causes an additional electrostatic confinement resulting in a small conducting island surrounded by a depleted region inside the constriction. Furthermore, the monotonic dependence of onset gate voltage versus constriction size (Figure~\ref{SizeVsGate}) and absence of island formation for constriction size of 370nm (Figure S1~\cite{SuppQD}) support this scenario in which larger (negative) gate voltage is needed to deplete the surroundings of the quantum dot formed.
\par
Another view is that the QD is formed due to the modified sample properties in a shallow region near the HSQ barrier. However, we demonstrate in Figure~\ref{Sample Characterization} that the \STO/\LAO~ remains intact in the vicinity of the HSQ and we demonstrate in Figures~\ref{SizeVsGate} and S1\cite{SuppQD} a monotonic dependence of the onset gate voltage on constriction size up to 370nm where the effect vanishes. Both these findings provide support to the electrostatic scenario we propose above.
\par
\subsection{Magnetic field dependence}
In Figure \ref{FanoField} we show the magnetic field dependence in two field orientations. The main important observation is absence of level splitting up to magnetic fields of 6T (see also Figures S4 and S5~\cite{SuppQD}). This suggests that the levels are not spin degenerate at the low carrier regime. For free electrons one expects a Zeeman splitting equivalent to a gate bias of 0.1Volt, well within the resolution of our measurements. Absence of spin degeneracy is also observed in ballistic transport in quantum wires \cite{ron2014one}.
\par
Our results are different than Cheng \etal~\cite{Levy2015QD}, where spin-splitting by a magnetic field is observed. It is possible that this is related to the difference between the \LAO~ layer below (Cheng \etal) and above (this contribution) the critical thickness \cite{kalisky2012critical}. We also note that magnetic effects have been related to the titanium $d_{xy}$ band, which is presumably the first to be populated in our quantum dot \cite{dxymagnetism2013lee}.
\par
Overall looking at the magnetic field dependence, in the perpendicular configuration the conductance background increases and the features are weakened (See Figure \ref{FanoField}(a) and Figure S5~\cite{SuppQD}). Sometimes a nonmonotonic behavior is observed, as shown in Figure \ref{FanoField}(b) and \ref{FanoField}(c) for perpendicular and transverse field orientations respectively. We attribute these observations to the magnetic flux threading between the interfering resonant and non-resonant paths.\cite{gores2000fano} Indeed, the above phenomena occur on a magnetic field scale corresponding to a flux quantum per square of size 20-30nm (the magnetic length), which is larger than the calculated island size (10nm), yet within the constriction (200nm for sample A1 and 140nm for sample B2). More work is still needed to clarify all the details of the magnetic field dependence in the different orientations.
\par
Strong spin orbit interaction could have resulted in an opposite effect namely, enhancement of coherent back-scattering with an applied magnetic field (similar to antilocalization). However, at the low carrier concentration regime in which the resonant and non-resonant channels interfere in our samples the spin-orbit interaction is strongly suppressed.\cite{caviglia2010tunable} We also note that in this low carrier regime superconductivity is fully suppressed.\cite{maniv2015strong}
\par
\section{CONCLUSION}
In summary, we developed a single-step lithography process suitable for conducting oxide interfaces. This process does not require any additional etching, lift-off or deposition of an amorphus layer. Using this process we were able to create a nanometric conducting island from a \STO/\LAO~interface. Conductance through this island exhibits features characteristic of a quantum dot, these features do not split at magnetic fields as high as 6T but they are suppressed by an additional parallel non-resonant current path. We analyzed our results within the framework of resonant channel interfering with a non-resonant one. From this analysis we find that the size of the island is of the order of 10nm. This size is much smaller than the lithographic size of the constriction. We conjecture that the geometry of sharp edges and the large nonlinear dielectric constant causes an additional electrostatic confinement. This lithography and electrostatic definition of the conducting regions can be used in future superconducting and electronic devices.
\par
\small
\begin{center}
\textbf{ACKNOWLEDGMENTS}
\end{center}
\normalsize
We thank Alon Kosloff for technical assistance and Eran Sela for useful discussions. This work was supported in part by the Israeli Science Foundation under grant number 569/13 by the Ministry of Science and Technology under contract 3-11875 and by the US-Israel bi-national science foundation (BSF) under grant 2014202. M.G. was supported by the Israeli Science Foundation under grant number 227/15.

\bibliographystyle{apsrev}
\bibliography{QD_bib}

\end{document}